\documentclass[final,12pt,sort&compress,times,3p]{elsarticle}

\usepackage{amssymb}
\usepackage{color}
\usepackage{url}
\usepackage{booktabs}

\journal{Physics Letters B}

\begin{document}

\begin{frontmatter}



\title{A possible correction of the Saha curve for non-equilibrium states}


\author[label1]{L. L. Sales}
\ead{lazarolima@uern.br}

\author[label1]{F. C. Carvalho}
\ead{fabiocabral@uern.br}

\author[label2]{H. T. C. M. Souza}
\ead{hidalyn.souza@ufersa.edu.br}

\affiliation[label1]{organization={Departamento de Física, Universidade do Estado do Rio Grande do Norte},
            city={Mossoró},
            postcode={59610-210}, 
            state={RN},
            country={Brazil}}

\affiliation[label2]{organization={Departamento de Ciências Exatas e Naturais, Universidade Federal Rural do Semi-Árido},
city={Pau dos Ferros},
postcode={59900-000}, 
state={RN},
country={Brazil}}

\begin{abstract}

It is widely known that the Saha equation is not suitable for describing plasmas out of thermodynamic equilibrium. The primordial hydrogen recombination plasma is an example of this. In this work, we propose a theoretical modification to the standard Saha curve motivated by Tsallis statistics. In particular, we explore the possibility that a time-dependent $q$-parameter may serve as an effective proxy for the evolving thermodynamic conditions during recombination, especially considering that hydrogen recombination occurs from excited states. Within this framework, the $q$-parameter could be interpreted as encoding departures from equilibrium and could play the role of effective time-dependent temperature. This indicates that the time evolution of the $q$-parameter could provide a phenomenological mechanism for incorporating non-equilibrium effects into the recombination history. Our findings suggest that the Tsallis parameterization provides an alternative path to fit the distribution of free electrons by using an effective temperature. The implications of this approach might go beyond its immediate applications, as the Saha equation is widely used in various scientific fields like astrophysics, cosmology, plasma physics, and condensed matter physics.

\end{abstract}




\end{frontmatter}


\section{Introduction} 

The Saha equation describes the ionization of atoms in plasmas under thermal equilibrium conditions \cite{Saha1920}. Within the Boltzmann–Gibbs (BG) statistical framework, it relates the free electron fraction to the temperature, and has been widely applied to problems in astrophysics and cosmology, from neutrino production in stellar interiors to primordial hydrogen recombination \cite{gangopadhyay2021hundred}. It also plays a role in Big Bang Nucleosynthesis (BBN), where it is used to estimate the abundances of light nuclei such as $^4$He. Despite its versatility, the Saha equation is strictly valid only for systems in thermodynamic equilibrium \cite{peebles2020principles,dodelson2024modern}, which limits its applicability to dynamical situations.

In the literature, it is well established that both BBN and primordial hydrogen recombination are more accurately described by kinetic approaches, such as the Boltzmann equation for annihilation \cite{peebles1968recombination,peebles2020principles,dodelson2024modern,zeldovich1969recombination}. This is because the relevant processes in the early Universe evolved from near-equilibrium to genuinely non-equilibrium conditions. As a result, the traditional Saha equation alone cannot provide a reliable description of hydrogen recombination.

Peebles argued that direct recombination to the hydrogen ground state produces photons capable of reionizing other atoms, yielding no net change. Instead, he suggested that recombination proceeds mainly through excited states \cite{peebles1968recombination}. This further undermines the applicability of the Saha equation, which assumes recombination directly to the ground state. More fundamentally, the limitation of Saha’s approach lies in its reliance on chemical equilibrium, which breaks down during certain stages of the Universe’s thermal history. Following Peebles’ pioneering work, successive studies refined the description of cosmological recombination by incorporating multi-level atom models, radiative transfer effects, and helium recombination, leading to more accurate predictions of the ionization history and the Cosmic Microwave Background anisotropies (see, e.g., \cite{seager1999new,dubrovich2005recombination,chluba2011towards,chluba2006free}).

The interactions in the early Universe are primarily determined by electromagnetic reactions, which exhibit long-range interactions. In 2018, J.L. Cirto et al. investigated the \textit{Validity and Failure of the Boltzmann Weight} in the context of long- and short-range interactions ($\alpha/d$) for potentials of the type $\Phi(r)\propto 1/r^{\alpha}$ ($\alpha\geq 0$) in a $d$-dimensional space \cite{cirto2018validity}. Accordingly, for short-range interactions ($\alpha/d>1$), numerical analysis indicates Maxwellians for the momenta and the Boltzmann weight for the energies. Nevertheless, for long-range interactions ($\alpha/d<1$), numerical analysis strongly suggests $q$-Gaussians for the time-averaged momenta, as well as $q$-exponential distributions for the Boltzmann weight, hence undoubtedly falling out of the scope of Boltzmann-Gibbs (BG) statistical mechanics. Since the Saha equation is proportional to the Boltzmann weight in the realm of BG statistics, it cannot accurately depict the physical phenomena in the presence of long-range interactions, such as electromagnetic interactions. Besides, the presence of long-range interactions or significant statistical correlations often drives the system away from equilibrium \cite{tirnakli2016standard}. Since the BG statistics is not well-suited to describe such non-equilibrium scenarios, modifications to the Saha equation become necessary to incorporate effects that lie beyond the BG paradigm.  

The Tsallis statistics is relevant in studying systems out of thermodynamic equilibrium that exhibit strong statistical correlations or long-range interactions \cite{tsallis2023introduction}. In particular, several studies have shown that space plasmas often display non-Maxwellian velocity distributions that can be accurately described by $q$-Gaussian distributions (also called kappa distributions), which are typical of the $S_q$-formalism \cite{livadiotis2013understanding}. This provides a robust physical motivation for exploring the role of non-extensive statistics in ionization–recombination processes. The limitations of the Saha equation have opened up a way to investigate its applicability in the early Universe under a generalized context. Motivated by this issue, we develop a strategy to correct the Saha ionization curve for primordial hydrogen recombination using Tsallis statistics \cite{tsallis1988possible}. Our main hypothesis is that a time-dependent $q$-index can act as an effective temperature, capturing the evolving thermodynamic conditions of the primordial plasma. This evolution is physically motivated by the fact that recombination occurs predominantly through excited states, a condition that the standard Saha equation does not contemplate. To this end, we shall employ the Saha ionization equation determined in Ref. \cite{sales2022non}. This modified version is distinct from the others because it does not include chemical potentials in its functional form, a common issue in alternative formulations (see \cite{pessah2001statistical,soares2019non}).

This work is structured as follows. In Sect. \ref{sect2}, an overview of primordial hydrogen recombination is presented. We discuss considerations on a parameterized model of recombination and a strategy to correct the Saha curve in Sect. \ref{sect3}. In Sect. \ref{sect4}, the results and discussion are shown. Finally, the conclusions are presented in Sect. \ref{sect5}. Unless otherwise specified, we will adopt the following system of natural units throughout the paper: $c=k_B=\hbar=1$.

\section{Primordial hydrogen recombination: an overview} \label{sect2}

In the context of cosmology, recombination is a process where electrons and protons combine to form neutral hydrogen atoms in the early Universe. This is one of the most significant transitions in the thermal history of the Universe, which took place at approximately 378000 years after the Big Bang. At the end of this era, nearly 99\% of the matter was combined with a significant amount of neutral hydrogen, with radiation temperature $T\sim 2500~{\rm K}$ \cite{peebles2020principles,peebles1968recombination}. The remaining radiation from the last scattering surface was first detected in 1965 by radio astronomers Penzias and Wilson \cite{penzias1965measurement}. This radiation is commonly called the Cosmic Microwave Background (CMB). 

In exploring the period of recombination in the early Universe, there are two distinct approaches for analysis: the Saha equation and the Boltzmann kinetic approach. Assuming that the reaction $e^{-}+p\leftrightarrow H+\gamma$ remains in equilibrium, the free electron fraction is calculated using the Saha equation \cite{dodelson2024modern}:
\begin{equation} \label{Saha-eq}
	\frac{x_{e}^{2}}{1-x_{e}} = \frac{1}{n_{b}\lambda_e^{3/2}}e^{-\varepsilon_{0}/T}~,
\end{equation}
where $\lambda_{e}=2\pi/(m_{e}T)$ is the thermal electron wavelength with $m_e$ being the electron rest mass, $\varepsilon_{0}=13.6~\textrm{eV}$ the ground state binding energy of the hydrogen atom and $T$ the thermal bath temperature. Also, $n_{b}=n_{e}+n_{H}$ is the baryon density calculated from the baryon-photon ratio by
\begin{equation}
	n_{b}=\eta_{b}\frac{2\zeta(3)}{\pi^2}T^3~,
\end{equation}
where $\eta_b=2.75 \times 10^{-8}\Omega_{b0}h^{2}$ with $\Omega_{b0}$ being the baryonic matter density parameter in the current time, $h$ is the reduced Hubble constant and $\zeta(x)$ the Riemann zeta function.

On the other hand, the numerical solution of the Boltzmann kinetic equation provides a more plausible description of recombination since it encompasses non-equilibrium states characteristic of that epoch of the Universe. For the sake of simplicity, we use the recombination model developed by Peebles \cite{peebles2020principles,peebles1968recombination}, which takes into account the expansion of the Universe, the photo-ionization due to the CMB, and the photons that arise from the recombination process:
\begin{equation} \label{EB-rec}
	\frac{dx_{e}}{dt} = \left[\beta_{e}(1-x_{e})e^{-[(B_{1}-B_{2})/k_{B}T]} - a_\mathrm{rec}x_{e}^{2}n_{b}\right]C \;.
\end{equation}
Here, $k_B$ is the Boltzmann constant, $B_{1}$ and $B_{2}$ are the binding energies of the ground state and the first excited state of the hydrogen atom, respectively. The photo-ionization rate reads as
\begin{equation} 
	\beta_{e} = a_\mathrm{rec}\left(\frac{m_{e}k_{B}T}{2\pi\hbar^{2}}\right)^{3/2}e^{-(B_{2}/k_{B}T)} \;,
\end{equation} 
in which $\hbar$ is the reduced Planck constant. The factor $C$ is the probability of an excited hydrogen atom decay via two photons emission ${2s}$-${1s}$, given by
\begin{equation} 
	C = \frac{\Lambda_{2s, 1s}}{\Lambda_{2s, 1s}+\beta_{e}} \;,
\end{equation}
where $\Lambda_{2s, 1s} = 8.23~\mathrm{s}^{-1}$ is the decay rate of two photons from the metastable level $2s$ to the ground state $1s$ \cite{peebles2020principles,jones1985ionisation}.

Besides, the recombination coefficient is given by
\begin{equation} 
	a_\mathrm{rec} \equiv \langle\sigma v\rangle = 2.84 \times 10^{-11}T_{m}^{-1/2}~\mathrm{cm}^{3}~\mathrm{s}^{-1} \;,
\end{equation} 
where $T_m$ is the matter temperature, ruled by the equation
\begin{eqnarray} \nonumber
	\frac{dT_m}{dt} &=& T_m\left[ -2H - \frac{\dot{x}_e}{3(1+x_e)}\right] -\frac{8}{3}\frac{\sigma_{T}bT^{4}x_e}{m_{e}c(1+x_e)}(T_m-T)~,
\end{eqnarray}
where $\sigma_{T}$ is the Thomson cross section, $c$ the speed of light in vacuum and $b=4\sigma/c$ with $\sigma$ being the Stefan-Boltzmann constant. In addition, $\dot{x}_e$ stands for the time derivative of the free electron fraction, and $H$ is the Hubble parameter. The cosmological redshift is related to radiation temperature by $T=T_0(1+z)$. In our calculations, we are assuming the flat $\Lambda$CDM model:
\begin{equation}
	\frac{H^2}{H_{0}^{2}} = \Omega_{m0}(1+z)^3 + \Omega_{r0}(1+z)^4 + \Omega_\Lambda~.
\end{equation}
For estimation purposes, we adopt the results from the Planck Space Telescope: $\Omega_{m0}=0.31$, $\Omega_\Lambda=0.68$, $\Omega_{r0}=5.47\times 10^{-5}$ and $H_0= 67.4~{\rm km}~{\rm s}^{-1}~{\rm Mpc}^{-1}$ \cite{aghanim2020planck}. 

The most accurate results for the ionization history are given by numerical codes, such as HYREC\footnote{Available at \url{https://github.com/nanoomlee/HYREC-2}.}, which take into account several important corrections to the cosmological recombination history \cite{ali2011hyrec,lee2020hyrec}. In the present study, we focus solely on the history of hydrogen recombination. We assume that helium recombination occurred well before this stage, ensuring that the ratios $n_e/n_b$ (Peebles approach) and $n_e/n_H$ (HYREC-2 code) are equivalent for $x_e<1$. We use the HYREC-2 in the full model, which is the default setting.

\begin{figure}[!ht]
	\centering
	\includegraphics[scale=0.7]{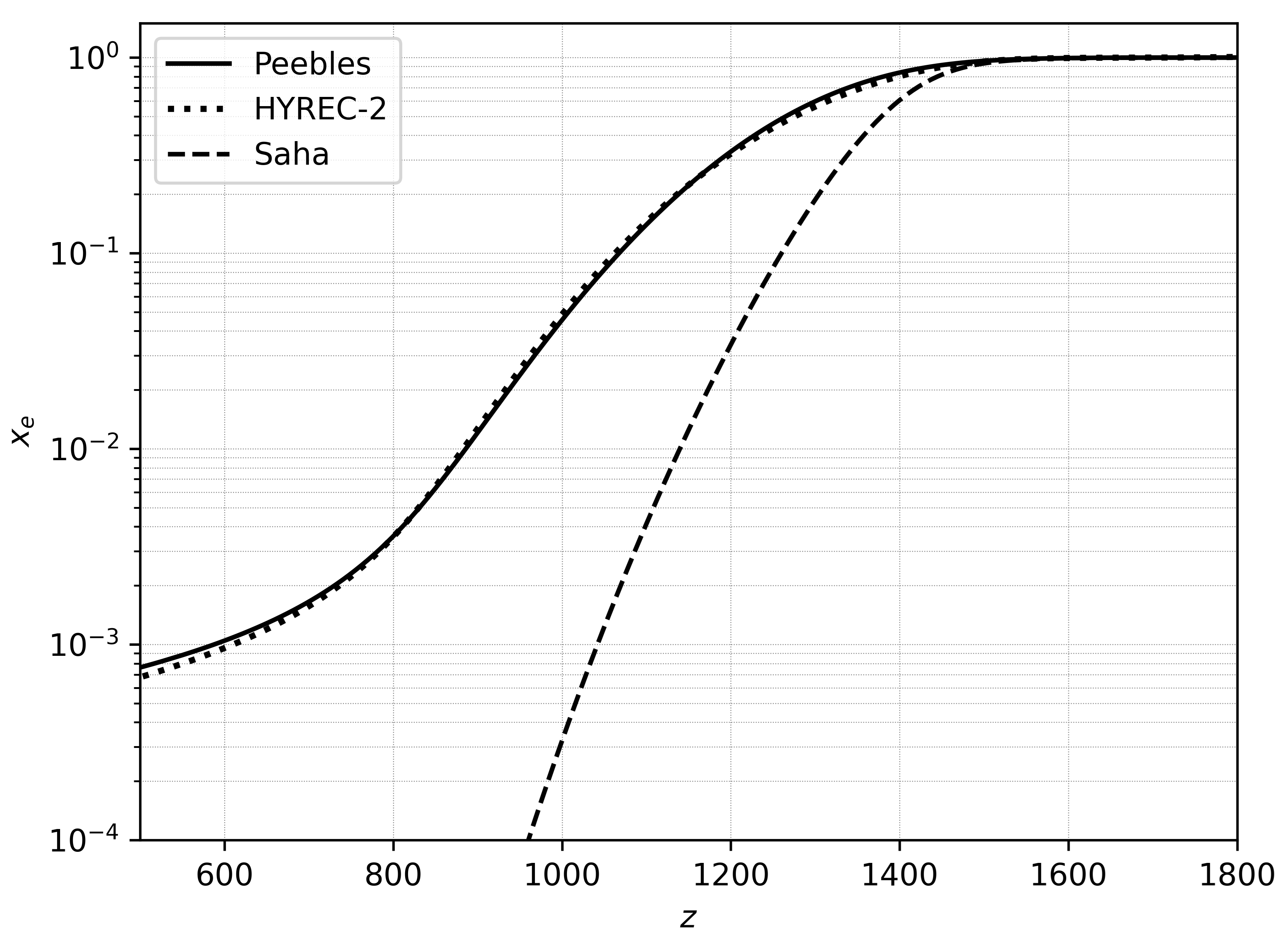}
	\caption{The curves depict the evolution of the ionization degree considering the Peebles (solid line) and Saha (dashed line) approaches and the result provided by the HYREC-2 code (dotted line).} 
	\label{fig.1}
\end{figure}

Fig. \ref{fig.1} shows the ionization history during recombination obtained through numerical integration of Eq. (\ref{EB-rec}), the HYREC-2 code, and the Saha equation. It is worth noting that the Peebles solution agrees reasonably well with the HYREC-2 code, while the Saha equation is only applicable as long as equilibrium is maintained. The HYREC-2 code and Peebles' approach predict a residual free electron fraction of about $\sim 10^{-4}$. This small fraction ionized at the end of recombination is important for the formation of the first hydrogen molecules in the Universe \cite{lepp1984molecules}.   

\section{A phenomenological model for hydrogen recombination} \label{sect3}

\subsection{Preliminary}

In 1988, C. Tsallis purposed a possible generalization of BG entropy \cite{tsallis1988possible}. The so-called $q$-entropy was defined as follows:
\begin{eqnarray}
	\label{qentropy}
	S_{q} = \frac{1}{q-1}\left(1-\sum_{i=1}^{W}p_{i}^{q}\right)~,
\end{eqnarray}
where $p_{i}$ is the normalized probability of the system in the microstate $i$, $W$ the total number of settings, and $q$ the parameter that measures the intensity of correlations of the system (deviation from standard equilibrium or the degree of non-additivity) commonly referred to as the entropic index. In the limit $q \rightarrow 1$, the BG entropy is recovered. 

In the Tsallis framework, the logarithm and exponential functions are redefined as
\begin{equation}
	\ln_q(x)=\frac{x^{1-q}-1}{1-q}~,
\end{equation}
and
\begin{equation}
	\exp_q(x)=e_{q}^{x}= {[1+(1-q) x]}^{1/{(1-q)}}~,
\end{equation}
respectively. These functions revert to their standard form when $q\rightarrow 1$.

The implementation of the $S_q$-formalism requires specifying how mean values are computed when maximizing the $q$-entropy. Following the classification proposed by Tsallis, Mendes, and Plastino (TMP) \cite{tsallis1998role}, three different definitions of internal energy $U_q$ can be employed. In this work, we adopt the second choice, which corresponds to the unnormalized expectation value,
	\begin{equation}
		U_q = \sum_{i} p_i^{q} E_i~,
	\end{equation}
	where $i$ stands for a given state with energy $E_{i}$. According to TMP, upon introducing a renormalized temperature, this formulation leads to the same stationary distribution as the one obtained with the normalized expectation (third choice of Ref. \cite{tsallis1998role}). Hence, both approaches can be regarded as effectively equivalent under such a renormalization procedure. Although the normalized expectation is often regarded as more rigorous from a thermodynamic standpoint, the unnormalized formalism remains consistent, widely used, and particularly suitable for phenomenological studies where the system departs from strict equilibrium conditions. In our analysis, this choice yields the non-Gaussian fermionic distribution, as discussed in Ref. \cite{sales2022non}. Interestingly, Mitra derived an identical fermionic distribution by adopting the third choice (normalized expectation) within the grand canonical ensemble, treating $T$ as the bulk temperature of the system \cite{mitra2018thermodynamics} (see also Refs. \cite{kikuchi2023reconsideration,ccimdiker2023equilibrium}).

Within the  $S_q$-formalism, we have shown in Ref \cite{sales2022non} that the particle number $q$-density for $q>1$ and $q<1$ reads as
\begin{equation} \label{q-density}
	n_{i}^{q} = \frac{g_{i}B_{q}}{\lambda_i^{3/2}}\left[ e_{q}^{\beta(\mu_{i}-m_{i})}\right]^{\frac{5-3q}{2}}~,
\end{equation}
where $\lambda_{i}=2\pi\beta/m_{i}$ (with $\beta=1/T$) and
\begin{eqnarray} 
	B_{q} = \left\{
	\begin{array}{rcl}
		\displaystyle{\frac{1}{(q-1)^{3/2}}\frac{\Gamma\left(\frac{5-3q}{2(q-1)}\right)}{\Gamma\left(\frac{1}{q-1}\right)}},& \mbox{if} & 1<q<5/3 \\ \\
		\displaystyle{\frac{1}{(1-q)^{3/2}}\frac{\Gamma\left(\frac{2-q}{1-q}\right)}{\Gamma\left(\frac{7-5q}{2(1-q)}\right)}}, & \mbox{if} & q<1~.
	\end{array}
	\right.
\end{eqnarray}
Moreover, $m_i$, $\mu_i$ and $g_i$ are the rest mass, chemical potential, and degeneracy of the species $i$, respectively.

The modified Saha ionization equation for the primordial hydrogen recombination is given by \cite{sales2022non}
\begin{equation} \label{q-saha-full}
	\frac{(x_{e}^{q})^2}{1-x_{e}^{q}} = \left(\frac{1.1\times 10^{16}}{\Omega_{b0}h^{2}}\right)D_{q}\left(\frac{T}{\rm eV}\right)^{-3/2}\left( e_{q}^{-\beta\varepsilon_{q}}\right)^{\frac{5-3q}{2}}\;,
\end{equation}
where 
\begin{equation} \label{gbe}
	\varepsilon_{q} =\frac{\varepsilon_{0}+(q-1)\beta m_{e}m_{p}}{1+(q-1)\beta m_{H}} 
\end{equation}
is the effective ionization potential and $D_{q} = B_{q}/C_{q}$ with
\begin{equation}
	C_q = \frac{2\psi^{(0)}(3-2q)-\psi^{(0)}(4-3q)-\psi^{(0)}(2-q)}{(q-1)^2}~,
\end{equation}
where $\psi^{(0)}(z)$ is the digamma function. It is worth stressing that Eq. (\ref{q-saha-full}) is constrained according to the range $0<q<4/3$.

\subsection{Strategy to correct the Saha curve}

In Ref. \cite{sales2022non}, it was shown that the $q$-parameter does not seem to be fixed throughout the recombination but should change with temperature. In this study, we assume that a time-dependent $q$-parameter acting as an effective temperature can mimic the evolving thermodynamic conditions during recombination in a way that reproduces the ionization history predicted by kinetic equations under non-equilibrium conditions. Besides, the effective ionization potential, Eq. (\ref{gbe}), suggests that recombination should occur from excited states for $q\neq 1$. This feature makes the generalized Saha equation a promising approach to correcting the Saha curve, since considering recombination to happen from the ground state is one of the main reasons why the standard Saha equation fails. Our strategy here is to apply the Peebles model and the HYREC-2 code to find a functional form for the $q$-parameter as a function of redshift in order to determine a parameterized recombination model (via the Saha equation).

Based on these premises, we employ a methodological procedure with a four-step strategy for correcting the Saha curve for primordial hydrogen recombination via the Tsallis framework. Each step is thoroughly described below:
\begin{enumerate}
	\item The first step deals with the solution of the Peebles equation for the degree of plasma ionization (free electron fraction $x_e$). Eq. (\ref{EB-rec}) is solved numerically, yielding a value of $x_e$ associated with each value of $z$. Also, we use the HYREC-2 code output;
	\item The second step is to rewrite the modified Saha ionization equation, Eq. (\ref{q-saha-full}), as follows:
	\begin{eqnarray} \label{q-z} 
		0 &=& \ln_q\left[ \left(\frac{x_{e}^{2}}{1-x_e}\right)\left(\frac{(1+z)^{3/2}}{1.364018\times 10^{23}D_q}\right)\right] ^{\frac{2}{5-3q}} + \frac{\varepsilon_{0}+(q-1)\frac{m_{e}m_{p}}{T_{0}(1+z)}}{T_{0}(1+z)+(q-1)m_{H}}~,
	\end{eqnarray}
	where $m_e=0.511~{\rm MeV}$, $m_p=938.272~{\rm MeV}$, $m_H=938.783~{\rm MeV}$ and $T_0=2.349\times 10^{-4}~{\rm eV}$. As $x_e$ is a function of redshift, the numerical solution of step 1 can be employed in Eq. (\ref{q-z}). This yields numerical solutions for $q$ at each $z$ value. Therefore, Eq. (\ref{q-z}) can be used to determine the appropriate variation of the $q$-parameter with the redshift. 
	\item Since the solutions obtained are numerical, it is not immediately clear if the functional form of the $q$-parameter can be determined in advance. As an {\it ad hoc} approach, we apply a polynomial fitting to determine a parameterization for the $q$-parameter from the numerical solution in step 2.
	\item In the final step, we incorporate a parameterization for $q(z)$ obtained in step 3 into the generalized Saha ionization equation, Eq. (\ref{q-saha-full}), resulting in a corrected version of the Saha equation for hydrogen recombination. Additionally, the exact numerical solution of step 2 can be used in Eq. (\ref{q-saha-full}) to ascertain if the curve generated overlaps the Peebles curve, which is expected.  
\end{enumerate}

\section{Results and discussion} \label{sect4}

\begin{figure}[ht]
	\centering
	\includegraphics[scale=0.7]{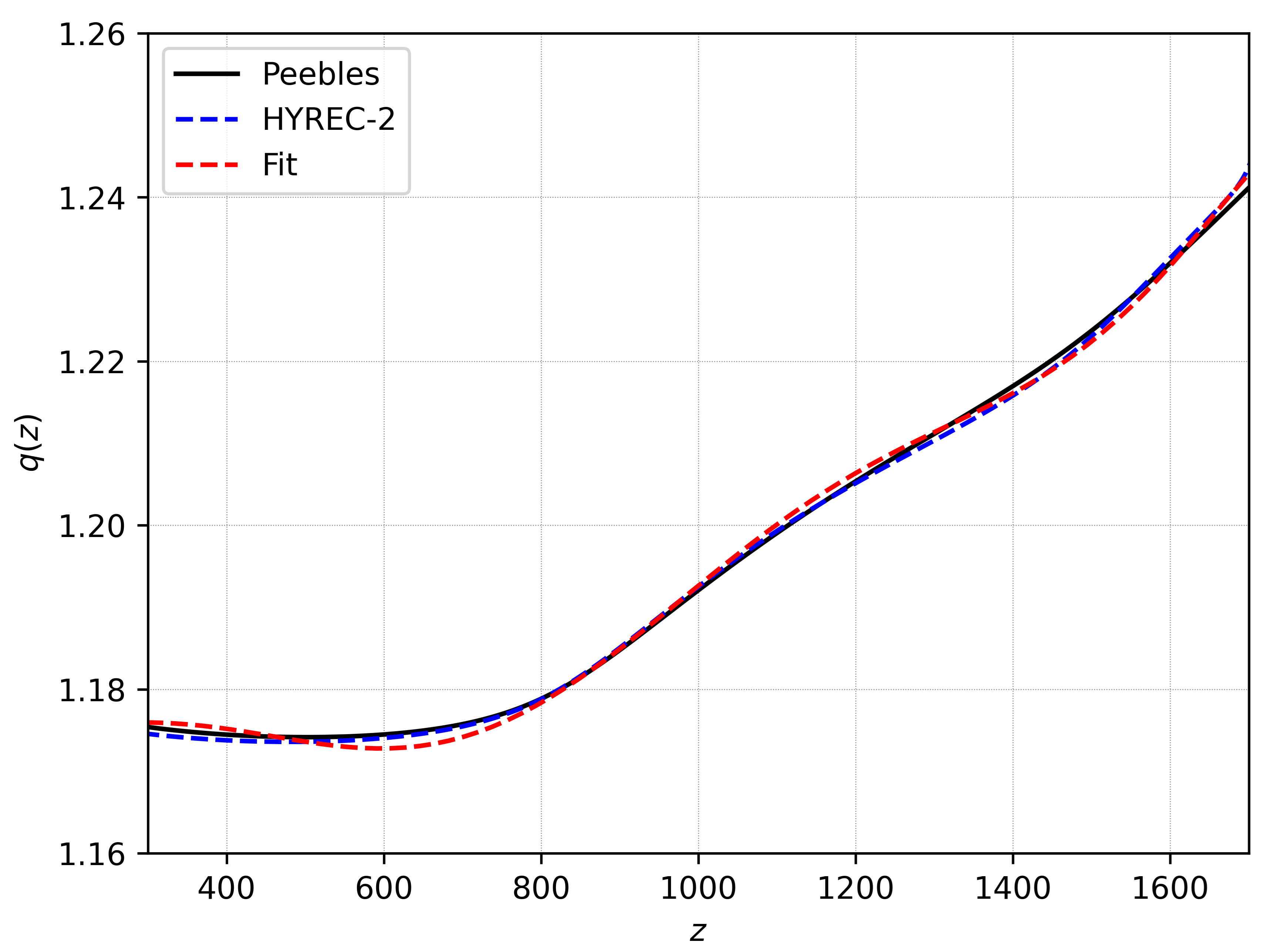}
	\caption{Numerical solution for the $q$-parameter as a function of redshift $z$ from step 2 of our strategy. The black solid line was obtained using the Peebles model. The blue and red dashed lines were yielded via the HYREC-2 code and seventh-degree polynomial fit, respectively.}
	\label{q-solution} 
\end{figure}

To correct the Saha curve, we will first focus on steps 2 and 3 of our strategy. Fig. \ref{q-solution} displays the graphical behavior of the $q$-parameter as a function of redshift, which was obtained from the numerical solution of Eq. (\ref{q-z}) using the Peebles model, the HYREC-2 code, and a polynomial fit. Fig. 2 illustrates that a constant $q$-parameter is unable to reproduce the ionization history, while the numerical solution clearly requires a time-dependent $q(z)$. Both the Peebles model and the HYREC-2 code provide consistent estimates for this variation, showing that departures from equilibrium can be effectively captured only if $q$ evolves with redshift. The polynomial fit is then introduced as a practical parameterization of this behavior, but the key physical result is that the correction to the Saha curve necessarily demands a non-constant $q$. 

For illustration purposes, we adopt a seventh-degree polynomial ($N=7$), whose coefficients are $a_0=1.201$, $a_1=-0.0003643$, $a_2=2.084\times 10^{-6}$, $a_3=-5.86\times 10^{-9}$, $a_4=8.585\times 10^{-12}$, $a_5=-6.623\times 10^{-15}$, $a_6=2.562\times 10^{-18}$, and $a_7=-3.919\times 10^{-22}$. This parameterization provides an approximate representation of the numerical solution for $q(z)$ and describes the ionization history reasonably well for $z \gtrsim 800$.

To evaluate the quality of the polynomial fit, we use standard statistical metrics: Mean Squared Error (MSE), Mean Absolute Error (MAE), Root Mean Squared Error (RMSE), and the coefficient of determination ($R^2$). For the comparison between the parameterized $q(z)$ and the Peebles model, we obtain ${\rm MSE}=0.000$, ${\rm MAE}=0.001$, ${\rm RMSE}=0.002$, and $R^2=0.986$. These values indicate that the polynomial fit reproduces the variation of the $q$-parameter with redshift during hydrogen recombination to reasonable accuracy. Similar results are found when comparing with the HYREC-2 solution.

Our approach enabled us to establish a minimum value for each model considered by lower bounding the $q$-parameter in the recombination epoch. For the models depicted in Fig. \ref{q-solution}, the minimum values found are: $q_{\rm Peebles}^{\rm min}\approx 1.1742$, $q_{\rm HYREC-2}^{\rm min}\approx 1.1736$, and $q_{\rm fit}^{\rm min}\approx 1.1728$, all fluctuating around $\sim 1.17$. Hence, any other recombination model must yield a minimum value of $q$ close to this range. Each of these minima corresponds to the point at which the system reaches the highest allowable entropy. Since HYREC-2 is a modern and accurate recombination code, $q_{\rm HYREC-2}^{\rm min}$ provides the most reliable estimate within our approach. 

Aside from the non-equilibrium aspect, the standard Saha equation fails by assuming that hydrogen recombination occurs directly from the ground state. Instead, more accurate models consider that recombination forms neutral atoms from higher excited states. In our study, we suggest that the temporal variation of the $q$-parameter is driven by its temperature dependence in excited states, i.e., $q \propto T$, as proposed in Ref. \cite{sales2022non}. More generally, recombination must proceed through excited states ($n>1$) because recombination directly to the ground state ($n=1$) yields $q = 1$, corresponding to the ordinary Saha result. Consequently, in a generalized non-equilibrium scenario, $q \neq 1$ is required, and the $q$-parameter must be temperature-dependent to account for recombination through excited states.

\begin{figure}[ht]
	\centering
	\includegraphics[scale=0.7]{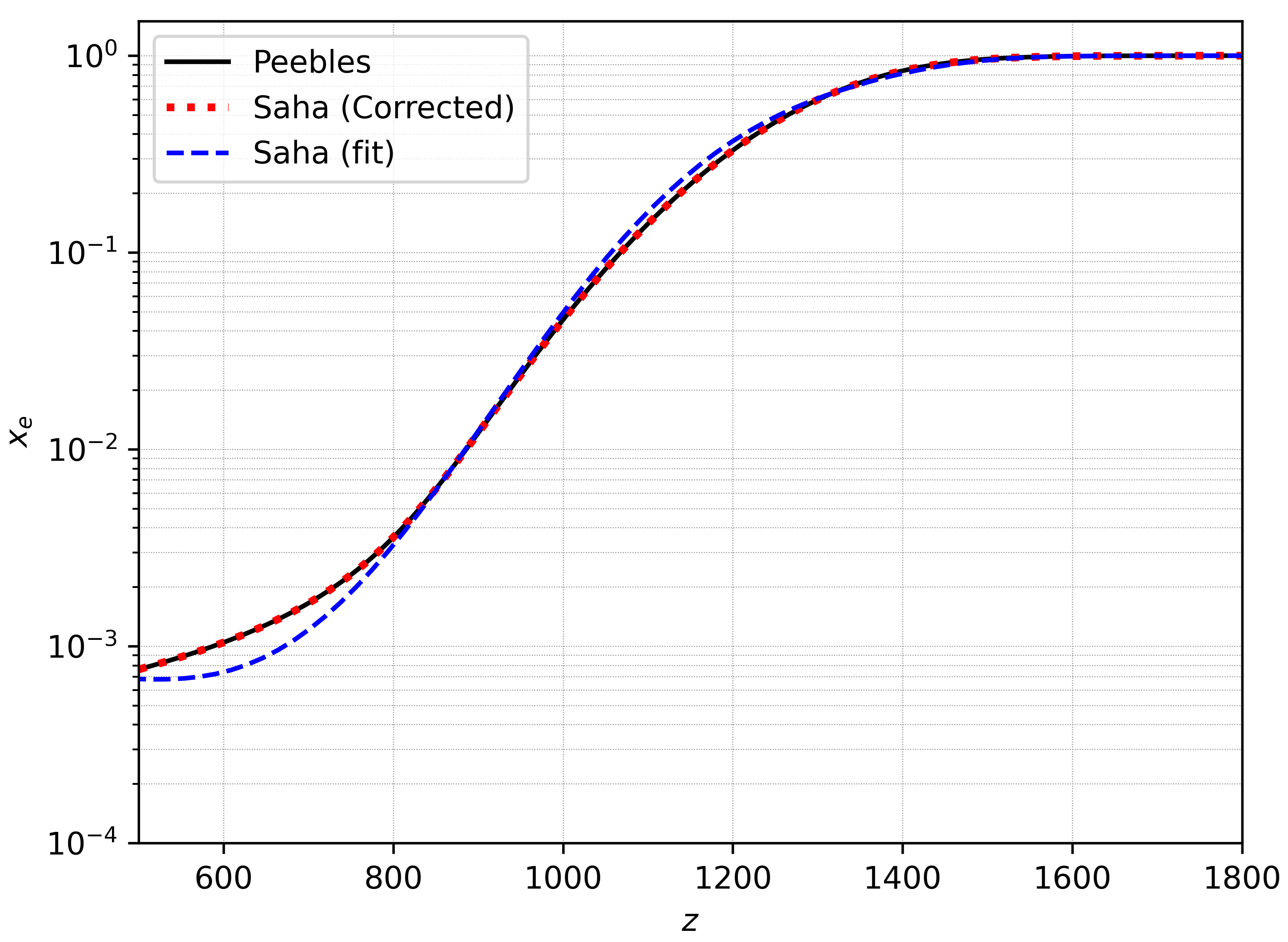}
	\caption{Evolution of the degree of ionization in the hydrogen recombination period. Three curves are shown: Peebles' approach (black solid line), Saha's ionization corrected (red dotted line), and Saha's 7th-degree polynomial fit (blue dashed line).}
	\label{q-rec}
\end{figure}

The variation of the $q$-parameter obtained from the degree-seven parameterization can be inserted into the generalized Saha equation. Since Eq. (\ref{q-saha-full}) applies only in the range $1<q<4/3$, the resulting evolution of the free electron fraction can be directly compared with the Peebles model. Figure \ref{q-rec} shows that the corrected Saha curve reproduces the ionization history reasonably well: the curve obtained with the fitted $q(z)$ (blue dashed line) closely follows the Peebles solution for $z \gtrsim 800$, with only small deviations. In addition, the red dashed curve in Fig. \ref{q-rec} corresponds to the numerical solution for $q(z)$ obtained directly from the Peebles model, providing a useful cross-check of our procedure. This demonstrates that a time-dependent $q$ is essential for bringing Saha’s approach closer to kinetic descriptions, a result that—to our knowledge—has not been explicitly reported before. For completeness, we also evaluated the quality of the fit using statistical metrics, obtaining ${\rm MSE}=0.0000$, ${\rm MAE}=0.0009$, ${\rm RMSE}=0.0045$, and $R^2=0.9994$. These values indicate that the polynomial parameterization reproduces the ionization history with good accuracy, although the essential point is that the correction requires a non-constant $q(z)$.

In general, the $q$-parameter is treated as a fixed quantity in most non-equilibrium physical systems (see, for example, \cite{hou2017non,tsallis2019beyond}). A constant value $q \neq 1$ describes a stationary non-Gaussian equilibrium state, where the system departs from the Boltzmann–Gibbs framework but still retains a well-defined maximum entropy. In this case, the effect of $q$ is essentially to shift the equilibrium point. By contrast, our results indicate that non-equilibrium situations such as primordial recombination can only be consistently captured if $q$ varies with time. When $q=q(t)$, the non-Gaussian equilibrium point itself evolves, reflecting the fact that the primordial plasma is subject to continuously changing thermodynamic conditions, particularly the role of excited states in the recombination process.

Our finding that the recombination history requires a time-dependent $q(z)$ is also consistent with independent arguments based on maximum entropy methods. Conroy and Miller \cite{conroy2015determining} showed that within the MAXENT framework, the Tsallis parameter may emerge as a function of the Lagrange multipliers associated with the system’s constraints, rather than as a fixed constant. In this case, $q$ evolves dynamically with the thermodynamic conditions. This provides theoretical support for our phenomenological result: during primordial recombination, the effective $q$ must vary with redshift, reflecting the evolving non-equilibrium state of the plasma rather than a static departure from Boltzmann–Gibbs statistics.

The interest in extending the nonextensive formalism to scenarios where the entropic parameter is not fixed was explored by Plastino, Miller, and Plastino in 2004 \cite{Plastino2004}. These authors developed a general thermostatistical framework in which both the entropic parameter and the quantities defining the relevant constraints may exhibit an effective temperature dependence, while preserving the Legendre-transform structure that connects the microscopic and macroscopic descriptions of thermodynamics. Specifically, they demonstrated that the extended variational formalism remains self-consistent when $q$ is allowed to vary, and that the stationary distributions retain the same formal properties as in the constant-$q$ case. Such developments provide a theoretical foundation for investigating systems in which the entropic parameter is not a universal constant but instead reflects the local or dynamical thermodynamic state. In this context, their approach may prove useful for describing cosmological plasmas with a time-dependent $q$, as considered in the present work.

It is worth stressing that the emergence of a minimum value for the $q$-parameter, around $q_{\min} \simeq 1.17$, acquires physical meaning. It marks the point of maximum allowed entropy during recombination, where the system is most strongly driven away from Boltzmann–Gibbs equilibrium while still being constrained by the non-Gaussian statistics. Since this minimum is consistently obtained whether one uses the Peebles model, HYREC-2, or the polynomial fit, it can be interpreted as a robust signature of the underlying thermodynamic evolution. In other words, the value of $q_{\min}$ does not depend on the details of the parameterization but reflects a physical property of the plasma during the recombination epoch.

\section{Conclusions} \label{sect5}

In this paper, we developed a strategy to correct the Saha curve using a generalized version of the Saha equation for non-equilibrium states. This approach is motivated by the fact that hydrogen recombination proceeds predominantly through excited states, a condition under which the standard Saha equation fails. Our central result is that the correction requires the entropic index $q$ to vary with time (or redshift). A constant $q \neq 1$ merely shifts the equilibrium point to a non-Gaussian stationary state, but does not reproduce the ionization history. By contrast, a time-dependent $q$ acts as an effective temperature, capturing the evolving thermodynamic conditions of the plasma.

From this perspective, the emergence of a minimum value $q_{\min} \simeq 1.17$ acquires physical meaning: it represents the point of maximum entropy allowed during recombination under non-Gaussian statistics. The fact that this minimum consistently arises from the Peebles model, HYREC-2, and the polynomial parameterization suggests that it is not an artifact of the fit, but rather a robust feature of the recombination dynamics. Thus, the temporal evolution of $q$ and the existence of a lower bound $\sim 1.17$ provide new physical insight into the role of non-extensive statistics in cosmology.

To illustrate the method, we adopted a seventh-degree polynomial to parametrize the numerical solution for $q(z)$, which reproduces the Peebles and HYREC-2 results with reasonable accuracy at $z \gtrsim 800$. This parameterization should be viewed as a practical tool rather than the main result: any function capable of capturing the required variation of $q(z)$—such as higher-order polynomials or non-parametric reconstructions—can be employed. What is essential is that the correction to the Saha curve demands a non-constant $q(z)$.

The broader relevance of this framework lies in its ability to reproduce well-established results in primordial cosmology from a novel statistical perspective. By replacing the assumption of equilibrium with a time-varying $q$, the generalized Saha equation becomes a physically motivated tool to describe ionization histories. This opens possibilities for applications beyond primordial recombination, such as modeling cosmic reionization, where parameterized forms of $x_e(z)$ are widely used, or in other areas where the Saha equation appears, including plasma physics and condensed matter systems.

Future work should explore the direct impact of a time-dependent $q(z)$ on cosmological observables, particularly the CMB, to test this framework against data. Such investigations may help establish whether the temporal behavior of $q$ can serve as a phenomenological proxy for non-equilibrium effects in the early Universe, further extending the physical scope of Tsallis statistics.

\section*{Acknowledgments}

FCC was supported by CNPq/FAPERN/PRONEM.  

  \bibliographystyle{elsarticle-num} 
  \bibliography{mybibfile}


\end{document}